\documentclass[12pt]{article}

\pdfoutput=1

\usepackage{amssymb,graphicx,slashed} 
\usepackage{epsf}
\usepackage{pstricks}
\usepackage{cite}

\newcommand{\beq}{\begin{eqnarray}}
\newcommand{\eeq}{\end{eqnarray}}

\newcommand{\centeron}[2]{{\setbox0=\hbox{#1}\setbox1=\hbox{#2}\ifdim

\wd1>\wd0\kern.5\wd1\kern-.5\wd0\fi \copy0

\kern-.5\wd0\kern-.5\wd1\copy1\ifdim\wd0>\wd1
                                       \kern.5\wd0\kern-.5\wd1\fi}}
\newcommand{\ltap}{\>\centeron{\raise.35ex\hbox{$<$}}
                               {\lower.65ex\hbox{$\sim$}}\>}
\newcommand{\gtap}{\>\centeron{\raise.35ex\hbox{$>$}}
                               {\lower.65ex\hbox{$\sim$}}\>}

\newcommand\ZZ{\hbox{\zfont Z\kern-.4emZ}}
\font\zfont = cmss10 

\begin{document}
\begin{titlepage}
\begin{flushright}
\end{flushright}

\vskip.5cm

\begin{center}
{\huge \bf Gaugephobic Higgs Signals}
\vskip.2cm
{\huge \bf at the LHC} 
\end{center}
\vskip.1cm
\begin{center}

\vskip.1cm
\end{center}
\vskip0.2cm

\begin{center}
{\bf Jamison Galloway$^a$, Bob McElrath$^b$, John McRaven$^a$, and John Terning$^a$}

\end{center}
\vskip 8pt

\begin{center}
$^{a}$ {\it Department of Physics, University of California, Davis,
CA
95616} \\
$^{b}$ {\it CERN, Geneva 23, Switzerland}\\
\vspace*{0.3cm}
{\tt  galloway@physics.ucdavis.edu, bob.mcelrath@cern.ch,
mcraven@physics.ucdavis.edu,  terning@physics.ucdavis.edu}
\end{center}

\vglue 0.3truecm

\begin{abstract}
\vskip 3pt \noindent
The Gaugephobic Higgs model provides an interpolation between three different models of electroweak symmetry breaking:  Higgsless models, Randall-Sundrum models, and the Standard Model.  At parameter points between the extremes, Standard Model Higgs signals are present at reduced rates, and Higgsless Kaluza-Klein excitations are present with shifted masses and couplings, as well as signals from exotic quarks necessary to protect the $Zb \bar b$ coupling.  Using a new implementation of the model in SHERPA, we show the LHC signals which differentiate the generic Gaugephobic Higgs model from its limiting cases.  These are all signals involving a Higgs coupling to a Kaluza-Klein gauge boson or quark.  We identify the clean signal $p p \to W^{(i)} \to W H$ mediated by a Kaluza-Klein $W$, which can be present at large rates and is enhanced for even Kaluza-Klein numbers.  Due to the very hard lepton coming from the $W^\pm$ decay, this signature has little background, and provides a better discovery channel for the Higgs than any of the Standard Model modes, over its entire mass range.  A Higgs radiated from new heavy quarks also has large rates, but is much less promising due to very high multiplicity final states.
\end{abstract}

\end{titlepage}



\section{Introduction}
\label{sec:Intro} \setcounter{equation}{0} \setcounter{footnote}{0}
A finite warped extra dimension has been known for several years as a candidate solution to the gauge hierarchy problem, as first shown in the Randall-Sundrum model, RS1 \cite{RS1}.  With two branes acting as the boundaries of the extra dimension, the geometry is such that gravity is sharply peaked at one end of the interval (towards the `UV brane'). The Standard Model (SM) fields residing at the other end (on the `IR brane') thus feel an exponentially diminished gravitational force at energies $\lesssim 1$ TeV.  Quantum corrections computed on a particular 4D slice of this spacetime will be cut off at a scale $\Lambda_z$ depending on position along the extra dimension $z$: if all SM fields are confined to the IR brane, then $\Lambda_{\rm SM}=\Lambda_{\rm IR} \sim 1$ TeV and the level of fine-tuning required to stabilize the weak scale is improved dramatically over the purely four-dimensional SM.

An instructive interpretation of this model can be gained from the AdS/CFT correspondence \cite{AdS/CFT}, which postulates relationships between the physics of the truncated extra dimension and the physics of a 4D approximately conformal field theory.  In particular, fields localized towards the IR brane in the fifth dimension are interpreted as composites of the spontaneously broken CFT, i.e. the scaling dimension of a 4D operator is directly related to the bulk mass of the corresponding 5D field.  In RS1, the confinement of all SM fields to the IR brane requires them each to have an infinite bulk mass, which implies an infinite scaling dimension in 4D.  This is clearly more than is required for a solution to the hierarchy problem: the only necessity is to push towards the IR brane any fields whose masses are not protected by a symmetry.  With this in mind we can naturally imagine more general models in which only the unprotected scalars of the theory (e.g. the Higgs) are strictly required to live in the vicinity of the IR brane.  

A realization of this sort of construction is provided by the Gaugephobic Higgs Model (GpHM) \cite{Gaugephobic}.  In this model all of the SM fields including the Higgs are free to propagate in the extra dimension.  The Higgs is forced via boundary conditions (BC's) to live near the IR brane such that the scaling dimension of the analog of the Higgs mass term, that is the square of the 4D operator associated with electroweak symmetry breaking (EWSB), is greater than four and is thus irrelevant.  
When the Higgs mass term is irrelevant then loop corrections are cutoff suppressed rather than divergent. Within this framework, one can interpolate between several models, including the 4D SM as well as Higgsless, holographic Technicolor, and composite Higgs models~\cite{NewCustodian,MCH,Higgsless}.

The GpHM presents a broad range of phenomenological possibilities.  Since electroweak symmetry is broken by a combination of BC's and a Higgs, generically all of the Higgs couplings to SM fields are suppressed.  This nullifies current bounds on the Higgs mass $m_H$, both experimental and theoretical, at both small and large mass.  There exists a large region of parameter space in which one identifies a suppressed $HZZ$ coupling responsible for the Higgs having avoided detection in the associated production channel at LEP \cite{LEP}.  At low masses, suppressed couplings to fermions can evade old limits, and provide a promising new search strategy at B-factories~\cite{Galloway:2008yh}.  The upper bound on the Higgs mass imposed by relying entirely on the Higgs for  unitarization of $WW$-scattering is modified, with new weakly coupled vector KK states contributing to that task~\cite{HiggslessUnitarity}. At the LHC, we might thus anticipate discovery of a very weakly coupled Higgs, or of no Higgs at all.  In each of these scenarios, the couplings of TeV-scale KK gauge bosons must be examined in order to determine if perturbative unitarity \cite{HiggslessUnitarity} is restored.~

In this paper we extend the study in ref. \cite{Gaugephobic} to determine the discovery possibilities at the LHC.   As the Higgsless model and Standard Model are limits of the GpHM, the LHC studies of those models are directly applicable, with the caveat that couplings may be suppressed and mass ranges widened with respect to those models \cite{Birkedal,Contino,Alves,Martin:2009gi}. Here we emphasize the importance of channels that are characteristic of the GpHM: channels including the participation of KK states {\it and} a Higgs.  Signatures of these channels can be used to distinguish the GpHM from the SM and traditional technicolor.

\section{The Gaugephobic Model}
\label{sec:Model} \setcounter{equation}{0} \setcounter{footnote}{0}
\subsection{Geometry and Field Content}
We work with the conformally flat metric:
\beq
ds^2 = \left(\frac{R}{z}\right)^2 \left(\eta_{\mu \nu} dx^\mu dx^\nu -dz^2\right).
\eeq
The extra dimension has boundaries at $z=R$ (the UV brane) and $z=R'\gg R$ (the IR brane).  With the coordinate $z$ corresponding to an inverse energy scale in the 4D theory, the gauge hierarchy problem can be tamed with appropriate choices of $R$ and  $R'$. The bulk gauge group is taken to be $SU(2)_L \times SU(2)_R \times U(1)_X$.     The $SU(2)_R$ is broken along with $U(1)_X$ at a high scale with Dirichlet BC's on the UV brane, i.e. $\left.SU(2)_R \times U(1)_X \right|_{z=R} \to U(1)_Y$.   The gauge fields have Neumann BC's on the IR brane since EWSB is triggered by the Higgs VEV.  This breaking preserves custodial isospin and leaves only $U(1)_{\rm EM}$ unbroken overall.  

With fermions introduced in doublets of the $SU(2)$'s, the $U(1)_X$ charge is simply related to the quantity $B - L$, though this arrangement turns out to be possible only for the first two generations.  If the top and bottom are arranged in doublets, the large overlap of the top with the IR brane required to achieve its high mass induces an experimentally unacceptable deviation in the coupling $g_{Zb_L \bar b_L}$.  We choose to resolve this problem by treating the third generation as in Ref.~\cite{NewCustodian}.  The group representations hosting the SM top and bottom quarks are chosen as
\beq
\Psi_L = ({\bf 2},{\bf 2})_{2/3} = \begin{pmatrix} {t_L & X_L \cr b_L & T_L} \end{pmatrix}; \qquad
\Psi_R = ({\bf 1},{\bf 3})_{2/3} = \begin{pmatrix} {X_R\cr T_R \cr b_R}\end{pmatrix}; \qquad
t_R = ({\bf 1},{\bf 1})_{2/3}.	
\eeq
With the right-handed $t$ introduced as a singlet, the top mass is fit without having to simultaneously move the $b$ so drastically  towards the IR as to induce a large shift in its coupling to the $Z$.  The new fields $T$ and $X$ have charges $2/3$ and $5/3$, respectively, so that $T$ mixes with $t$ to form mass eigenstates.  We show in Sec.~\ref{sec:WWH} the enhancing effect that some of the new quarks have on processes involving other SM fields.

The Higgs is  a bidoublet of $SU(2)_L \times SU(2)_R$ with zero $X$ charge.  The Higgs VEV is induced by a quartic interaction which is confined to the IR brane to ensure that EWSB takes place at the appropriate energy scale.  The diagonal VEV breaks $SU(2)_L \times SU(2)_R \rightarrow SU(2)_{\rm diag}$, which maintains a custodial symmetry as usual. A mass term for the Higgs on the UV brane then selects a profile for the VEV of the form
\beq
v(z) = a \left(\frac{z}{R}\right)^{2 +\beta}.
\eeq
Here $\beta$ is a convenient parameterization of the bulk mass $\mu$ defined by $\beta = \sqrt{4 +\mu^2}$; we see then that $\beta$ dictates the localization of $v$.  This parameterization makes clear the fact that the scaling  dimension of the 4D CFT operator responsible for breaking electroweak symmetry is $2+\beta$.  Choosing a value $\beta > 0$,  the VEV gets pushed towards the IR, which is equivalent in 4D language to the square of the CFT operator responsible for EWSB becoming irrelevant.  A normalization, $V$, is chosen \cite{Gaugephobic} so as to write 
\beq\label{eq:VEVProfile}
v(z) = \sqrt{ \frac{2 (1+\beta) \log R'/R}{1-(R/R')^{2+2\beta}}} \frac{g V}{g_5} \frac{R'}{R} \left(\frac{z}{R'}\right)^{2+\beta}
\eeq
with $g$ the SM $SU(2)$ coupling and $g_5$ the coupling of the bulk $SU(2)$'s.

\subsection{Limiting Cases and Benchmarks}

The GpHM encompasses a large class of models.  In particular limits of the parameter space, we can match exactly onto the SM, RS1\cite{RS1}, Higgsless models \cite{Higgsless} and composite Higgs models as in \cite{MCH}.  These limits can all be obtained by varying the three parameters describing the profile of the Higgs VEV, shown in Table~\ref{table:Models}.  We find it convenient to describe this parameter space with the bulk mass $\beta$, the profile's normalization factor $V$, and the physical Higgs mass $m_H$ (keeping the gauge boson masses and couplings fixed).  In this case, the purely 4D SM is recovered by setting $V=246 \, {\rm GeV}$, and $\beta=-1$.  As $V$ is increased, the deviation from the SM becomes apparent as all Higgs couplings are decreased.  Traditional RS with a composite Higgs is obtained by pushing the Higgs all the way to the IR brane by taking $\beta \to \infty$.  We approach the Higgsless limit by  pushing the Higgs to the IR and taking $V \to \infty$.  In this limit, the BC's in the IR for the gauge bosons switch to Dirichlet and all mass  comes entirely from momentum in the extra dimension.  

In this study we explore further the two illustrative benchmark points defined in \cite{Gaugephobic}.  In both cases we take $\beta >0$ to solve the gauge hierarchy problem; the variation between the two points is  entirely in $V$. The first, which we refer to as the {\it gaugephobic scenario}, sets $V=300 \, {\rm GeV}$ and $\beta =2$.  The $ZZH$ coupling is suppressed relative to its SM value by about a factor of 2 in this case.  This can explain the nonappearance of a light Higgs at LEP, but still allows for discovery at the LHC.  We'll refer to the second benchmark as the {\it Higgsless scenario}, and sets $V=500 \, {\rm GeV}$ and $\beta =2$.  Here $ZZH$ is suppressed by a factor of about 10, and the world will appear Higgsless to experiments performed at the LHC.  
\begin{table}[ht]
    \begin{center}
\begin{tabular}{l|cc}
                                  & $\beta$ & V  \\
    \hline
    Higgsless~\cite{Higgsless}    &     $\infty$     & $\infty$  \\
     RS1 ~\cite{RS1} & $\infty$  & $\sim$ 246 GeV \\
    Mixed Higgsless benchmark~\cite{Gaugephobic}
    & 2        & 500 GeV   \\
    Mixed gaugephobic benchmark~\cite{Gaugephobic}
    & 2        & 300 GeV   \\
    RS Composite Higgs~\cite{MCH} & 0 & $\sim$ 246 GeV \\
    Unhiggs~\cite{unhiggs} & $\sim$ -0.3 &   $\sim$ 246 GeV\\
    Standard Model                &      -1    & 246 GeV   
\end{tabular}
\end{center}
\caption{Gaugephobic parameters for benchmark points studied, and limits for other models.}\label{table:Models}
\end{table}

\subsection{Massive Vector Bosons: Spectrum and Couplings}
All SM fields are in the bulk, so each has its own KK tower.  The presence of neutral KK gauge bosons is particularly straightforward to ascertain due to their characteristic dilepton final states; we show this concretely for the two benchmark points in the next section.   Discovery of a heavy neutral gauge boson alone, however, would still leave open different interpretations: several models---including some 4D constructions---predict such states.  To disentangle the possibilities, more detailed information must be sought. 

\begin{table}[ht]
\begin{center}
\begin{tabular}{c|c|c|c|c}
State & Mass & $g_{W^{(i)}u_Ld_L}/g_{Wu_Ld_L({\rm SM})}$ &  $g_{W^{(i)}WH}$/$g_{WWH({\rm SM})}$ & $g_{W^{(i)}W^{(i)}H}$/$g_{WWH({\rm SM})}$ \\
\hline \hline
$W^{(1)}$ & 918 &  0.15  & 0.56 & 0.595 \\
$W^{(2)}$ & 1114 & 0.25  & 4.79 & 44.56 \\
$W^{(3)}$ & 2075 & 0.07  & 1.18 &  3.93 \\
$W^{(4)}$ & 2164 & 0.10 & 3.56 &  37.35  \\ \hline 
\end{tabular}\caption{Spectrum and couplings of first KK gauge bosons in GpHM at $V=300\, {\rm GeV}.$}\label{table:W}
\end{center}
\end{table}

It turns out, as we'll see in Sec.~\ref{sec:WWH}, that the easiest way to distinguish the GpHM from other extensions of the SM is through channels involving KK excitations of the charged gauge bosons.  The most striking characteristic is the coupling of KK $W$'s to the Higgs; generically we find that the Higgs couples preferentially to the $W^{(2n)}$ states as shown in Table~\ref{table:W}.  This preference for even KK states can be understood as a result of the form of the Higgs couplings in the 5D Lagrangian.  The important point is that the 5D Higgs couples to the weak gauge bosons with a strength that depends on the {\it difference} between the portion of the gauge boson profile coming from the bulk $SU(2)_L$ and that coming from the $SU(2)_R$.  More precisely, the coupling $g_{W^{(i)} W^{(j)} H}$ comes from integrating the function of profiles
\beq
g (z) = \frac{1}{2} v\, h \left(W_L^+ - W_R^+\right)^{(i)}_\mu \left( W_L^- - W_R^-\right)^{\mu, (j)}
\eeq
over the extra dimension.  The BC's for the $W$ fields are such that in the vicinity of the IR brane, the odd KK states have $W_L \sim W_R$ while even states have $W_L \sim -W_R$.  In the Higgsless limit, for instance, the BC's at the IR brane are exactly\footnote{The modifications to these BC's coming from the existence of the bulk Higgs in the GpHM result in profiles that are qualitatively similar enough to allow comparison between the two cases.}
\beq
W_L &=& W_R \\
\partial_z W_L &=& -\partial_z W_R \, ;
\eeq
the first $W$ excitation is nearly flat in the IR yielding $W_L \sim W_R$ near the brane, while the second mode has non-zero slope and thus $W_L$ and $W_R$ diverge away from the brane.  Note that this indicates that any coupling enhancement is diminished by increasing the Higgs bulk mass (which pushes the Higgs into a region where the profiles of the separate KK $W$'s become increasingly similar) or by increasing $V$.  When both of these parameters are taken to be large we thus recover the anticipated Higgslessness.

\section{Standard Model Search Channels}
\label{sec:SM} \setcounter{equation}{0} \setcounter{footnote}{0}
The GpHM will have discernible effects in orthodox search channels at hadron colliders.  As all Higgs couplings to SM fields are suppressed, we have the obvious expectation of decreased statistical significance in most standard search modes.  Before moving on to discuss strategies for more unique identification of the model, we show a representative case of the statistical impact that a gaugephobic Higgs would have in processes investigated at the LHC.  We discuss also the required appearance of KK states for the gauge bosons.  The contribution of a $Z'$ to di-muon final states, for instance, would be apparent early in the running of the LHC.

\subsection{Standard Higgs Search at the Gaugephobic Benchmark}
The Higgs production/decay rate is suppressed in the entire parameter space.  Since ratios of couplings remain similar to the SM case, however, we can meaningfully explore the usual channels for Higgs searches in the gaugephobic scenario where $g_{ZZH}$ reaches $\sim 10\%$ of its SM value.  In particular, the Higgs is produced predominantly via vector boson fusion (VBF) and follows typical decay patterns.  Here we study the case where the Higgs decays via $H \to ZZ^* \to 4\mu$.  For a Higgs mass around 160 GeV, the $ZZ$ and $WW$ channels dominate (as in the SM) so that from the former we anticipate a sufficient number of four-muon final states to easily resolve the Higgs.  Overall we expect two hard forward jets and four muons of invariant mass $\sim m_H$ appearing in this channel.  

\begin{figure}[t]
\begin{center}
\includegraphics[width=11cm]{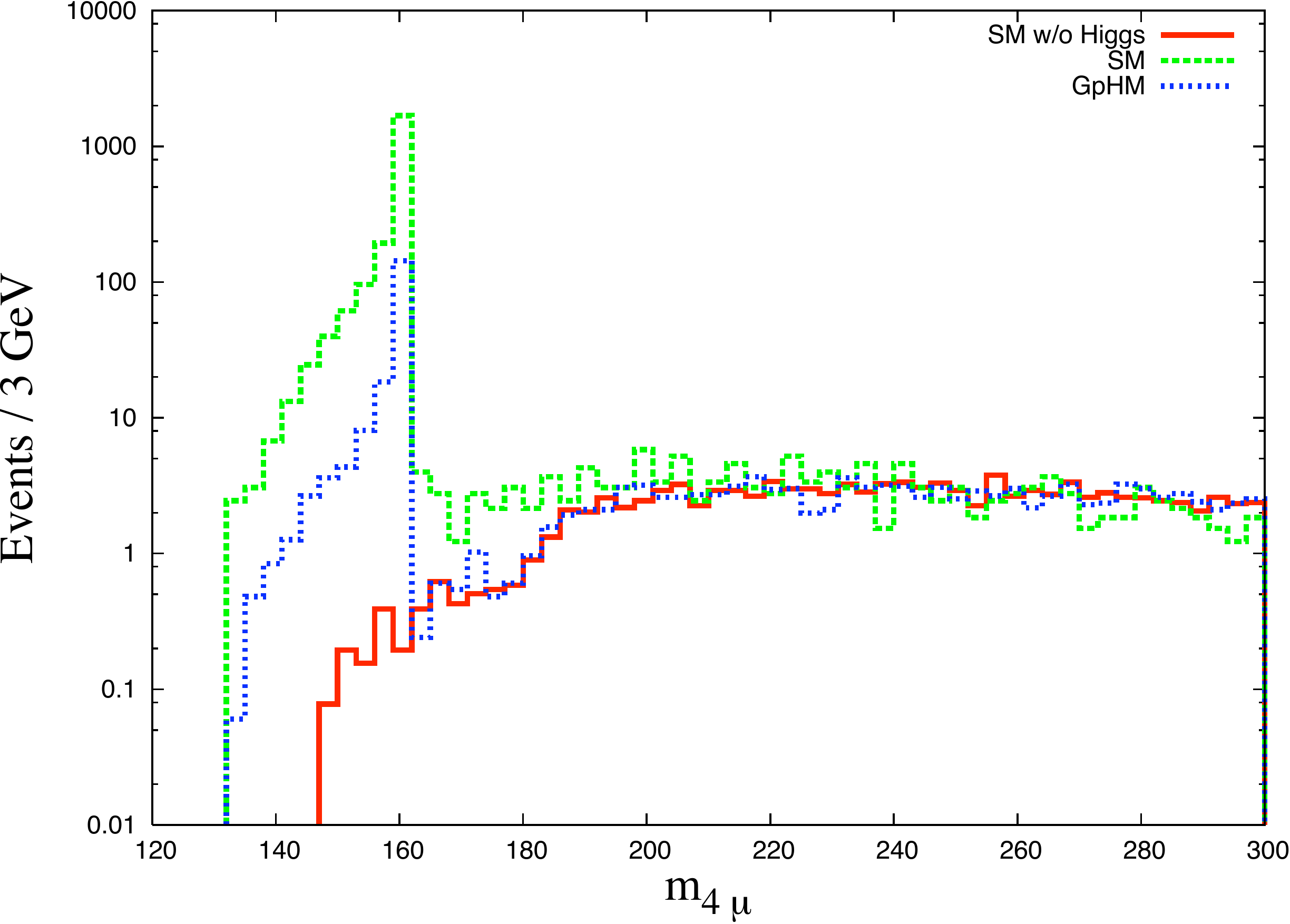}
\end{center}
\caption{Distribution of the final state from  $pp \to 4 \mu + 2j$ with an intermediate gaugephobic Higgs at $m=160\, {\rm GeV}$ assuming an integrated luminosity of 300 ${\rm fb}^{-1}$.  In the gaugephobic Higgs model we find a simple scaling of the SM result coming from the characteristic decreased couplings.}\label{fig:GPHiggs}
\end{figure}

The couplings used in this study are determined numerically by integration of profile overlaps over the extent of the extra dimension.  These couplings are read into CompHEP \cite{CompHEP} to compute widths of all unstable particles, and event generation is completed with SHERPA \cite{SHERPA}.  Where possible, cross-sections have been checked analytically, and SHERPA output in simple cases has been confirmed to closely match predictions from PYTHIA \cite{PYTHIA}.  For the  channel at hand, the requirements imposed on the jets are implemented in the event generation with the following $p_T$ and isolation cuts:
\beq
p_{Tj}>20\; {\rm GeV},  \quad \Delta R_{jj}>0.6.
\eeq
Approximating detector acceptance amounts to a cut on the pseudo-rapidities of the jets and muons; we take $|\eta_j|<5, \ |\eta_\mu|<2.5$.
We also stipulate that two of the muons can be identified with the on-shell $Z$ decay by requiring $|m_{\mu_1  \mu_1 } - m_Z| < 15$ GeV.  Finally, to isolate events coming from weak boson fusion we require 
\beq
m_{jj} > 600, \quad |\Delta \eta_{j j}| > 4.2, \quad \eta_{j_1} \cdot \eta_{j_2} < 0.
\eeq
Low energy events are excluded with a requirement on the second muon pair, $m_{\mu_2  \mu_2} > 40 \; {\rm GeV}$.  The distribution of this final state with these cuts is shown in Fig.~\ref{fig:GPHiggs}.  We assume an integrated luminosity at the LHC of 300 ${\rm fb}^{-1}$ throughout.

Fig.~\ref{fig:GPHiggs} shows the anticipated scaling of SM predictions.  This sort of scaling will be found in all cases where processes are dominated by contributions from SM fields.   Thus a light ($\sim 120 \, {\rm GeV}$) gaugephobic Higgs would go undetected in the majority of the parameter space if only standard channels are relied upon.  Analysis of the standard preferred decay channels for this mass range does not afford any new insight or amendment to this conclusion.  We'll see in Sec.~\ref{sec:WWH}, however, the utility of new channels in the GpHM as probes for a light Higgs.

\subsection{$Z'$ Resonances in $pp \to \mu \mu$}
\begin{figure}[t]
\begin{center}
\includegraphics[width=11cm]{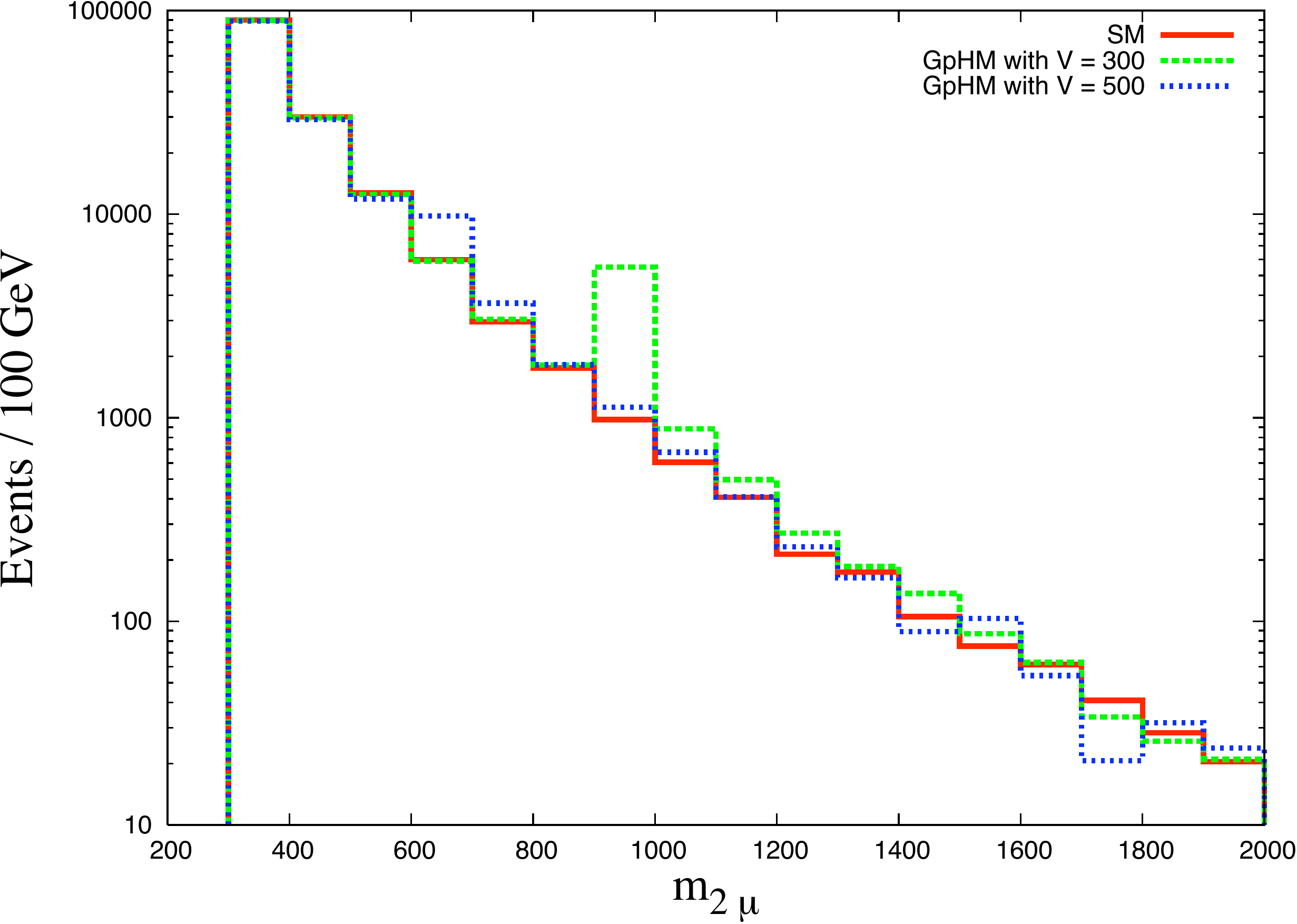}
\end{center}
\caption{Distribution of $2 \mu$ final state, assuming an integrated luminosity of 300 ${\rm fb}^{-1}$, for the gaugephobic and Higgsless benchmarks.  One can clearly see the combined effect of the neutral KK gauge bosons at 914 GeV and 944 GeV when $V=300$; and 597 GeV and 617 GeV when $V=500$.  The excitation with smaller mass contributes significantly less in each case.
}
\label{fig:Z'}
\end{figure}

With the suppressed Higgs couplings comes the requirement of KK gauge bosons entering with $m\sim 1$ TeV, and we can easily search for such states arising as resonances in a Drell-Yan process.   These resonances would also appear in $WW$ scattering; a brief discussion of this channel is given below.  At high $p_T$, jet effects will not significantly affect the signal of such a resonance decaying to a lepton pair, so in the simulation we examine a simple $\mu^+ \mu^-$ final state ignoring jets.  Generated data for this final state at both benchmark points is shown  in Fig.~\ref{fig:Z'}.  Note that we have contributions coming from two neutral KK states at each benchmark: the masses of the first KK excitations of the $Z$ and $\gamma$ are too similar to be individually resolved so appear as a single resonance.  

The second benchmark point is chosen as a representative of regions of parameter space where the Higgs is decoupled as far as the LHC is concerned.  In this scenario, there are several non-trivial sum rules that should be satisfied between the KK gauge bosons' couplings for restoration of unitarity \cite{HiggslessUnitarity}, thus allowing for a more definitive verification of the model.  This issue has been examined within a study of Higgsless models \cite{Birkedal}, where for validation of the sum rules an intermediate massive vector boson (MVB) is produced from the fusion of two SM gauge bosons originating from Bremsstrahlung off of two incoming partons.  The direct coupling of the new MVB's to the SM gauge bosons sets apart scenarios like this from other SM extensions where the new vectors come from entirely new gauge groups (cf. reviews in \cite{Rizzo,Langacker}); it is because of this distinguishing characteristic that VBF production is preferred in these extra-dimensional models.    Furthermore, this channel allows for a more model-independent study, but complicates $Z'$ searches due to large backgrounds \cite{WWLHC} and the required reconstruction of hadronincally decaying $W$'s.   Here again we focus on a specific $Z'$ whose discovery would be well within the reach of the LHC, affording a simple viability check of this model which could be valuable in the absence of a Higgs discovery.  We see in Fig.~\ref{fig:Z'} that the expected distribution is qualitatively similar to the $Z'$ anticipated at the gaugephobic benchmark, with the requisite decrease in mass.

\section{Strategies for a Gaugephobic Higgs}
\label{sec:WWH} \setcounter{equation}{0} \setcounter{footnote}{0}
\subsection{$W'$ Higgsstrahlung}

We turn now to some qualitatively new Higgs search possibilities.   In the SM, a Higgs can be produced simply by radiation off of a $W$ or $Z$.  This is not a promising channel for a light Higgs due to irreducible backgrounds, but in the Gaugephobic scenario we're afforded a new possibility where the Higgs is radiated from an excited state, e.g. a $W'$ produced via Drell-Yan.  This channel exists in other models as well, e.g. Little Higgs (see for instance \cite{LittleHiggs}), but we find a distinguishing situation here due to the strengthened couplings between the Higgs and the KK gauge bosons.  A heavy $W$ decaying leptonically gives a clean single isolated lepton final state to trigger on and provides a possible ``Golden Channel" if the coupling of the $W$ to other particles is significant.  

\begin{figure}[ht]
\centerline{\includegraphics[width=11cm]{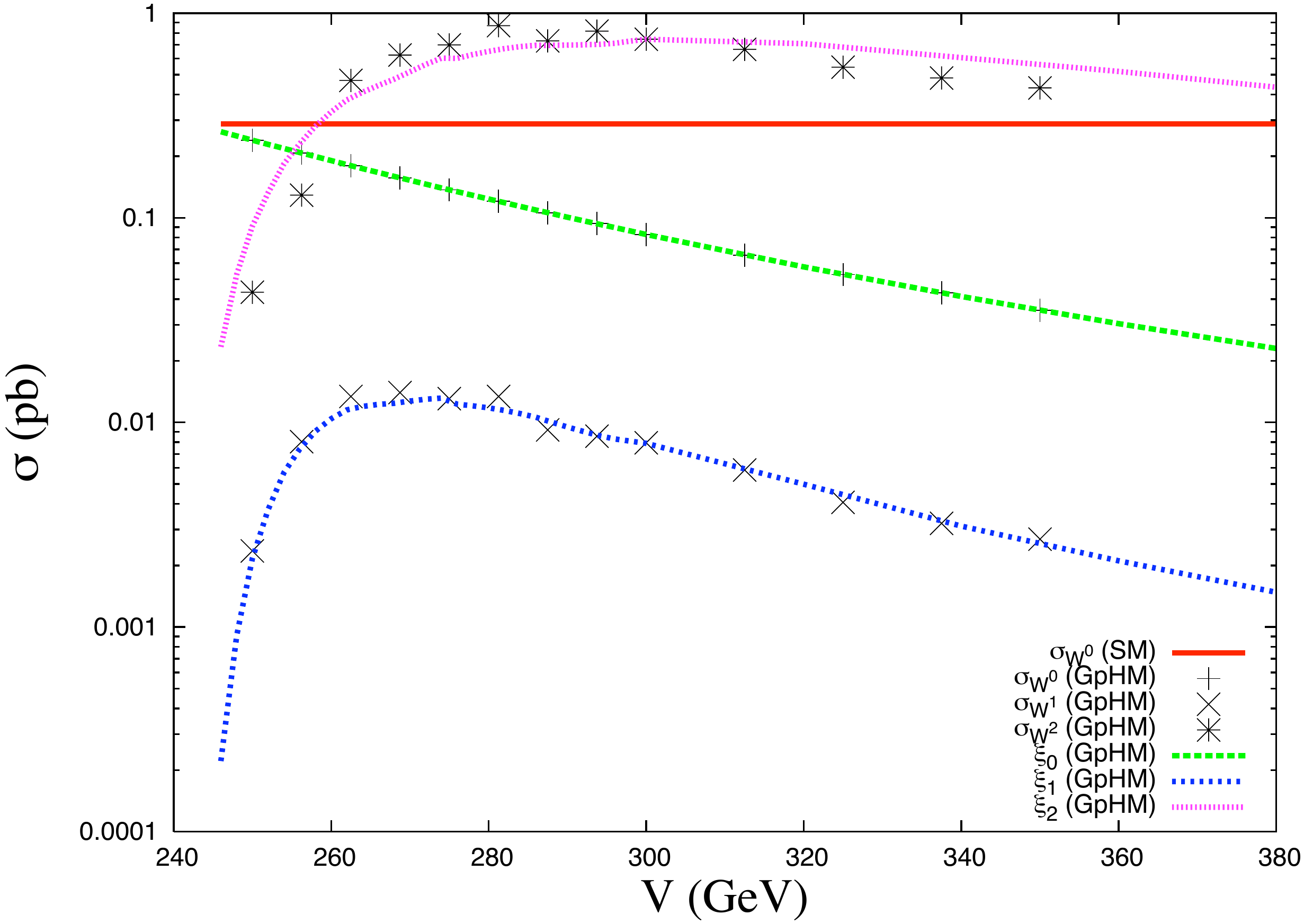}}
\caption{Approximate contribution to $pp \to W H$ from an intermediate $W'$; see Eq.~(\ref{eq:xi}).  }
\label{fig:xi_WH}
\end{figure}

We have shown in Sec.~\ref{sec:Model} that the second $W'$ state has the most favorable coupling to the Higgs.  The values quoted there have a dependence on the profile of the Higgs VEV: as $V$ increases, we expect Higgs to $W$ couplings to decrease provided masses are held constant (just as in the SM where $g_{WWH} \sim m_W^2/v$).  It is natural to wonder then how useful this channel will be at other points in the parameter space.  In Fig.~\ref{fig:xi_WH} we have examined a quantity we define as
\beq\label{eq:xi}
\xi_{i} = \frac{g^2_{W^{(i)}ud} \times g^2_{W^{(i)} W H}}{m_{W^{(i)}}^4/m_{W^{(0)}}^4}
\eeq
(with all couplings defined relative to their SM counterparts)  as a function of $V$.  This effective coupling, appropriately scaled, encodes practically all $V$-dependence of the cross-section.  Only the contribution from the first generation of quarks in the proton is considered.  $\xi_0$ is scaled so that it approaches the SM cross section in the proper limit.  $\xi_1$ and $\xi_2$ are scaled such that $\xi_2$ matches the computed value of $\sigma(pp\to W^{(2)} \to W H)$ when $V=300$.
We have confirmed the validity of this approximation with full computations of $\sigma(pp \to WH)$ at several points with varying $V$.  These are shown also in Fig.~\ref{fig:xi_WH}.  We find that although the coupling enhancement is nearly maximal at the gaugephobic benchmark, its effect will be apparent in the neighborhood of parameter space surrounding that point.

\begin{figure}[ht]
\centerline{\includegraphics[width=11cm]{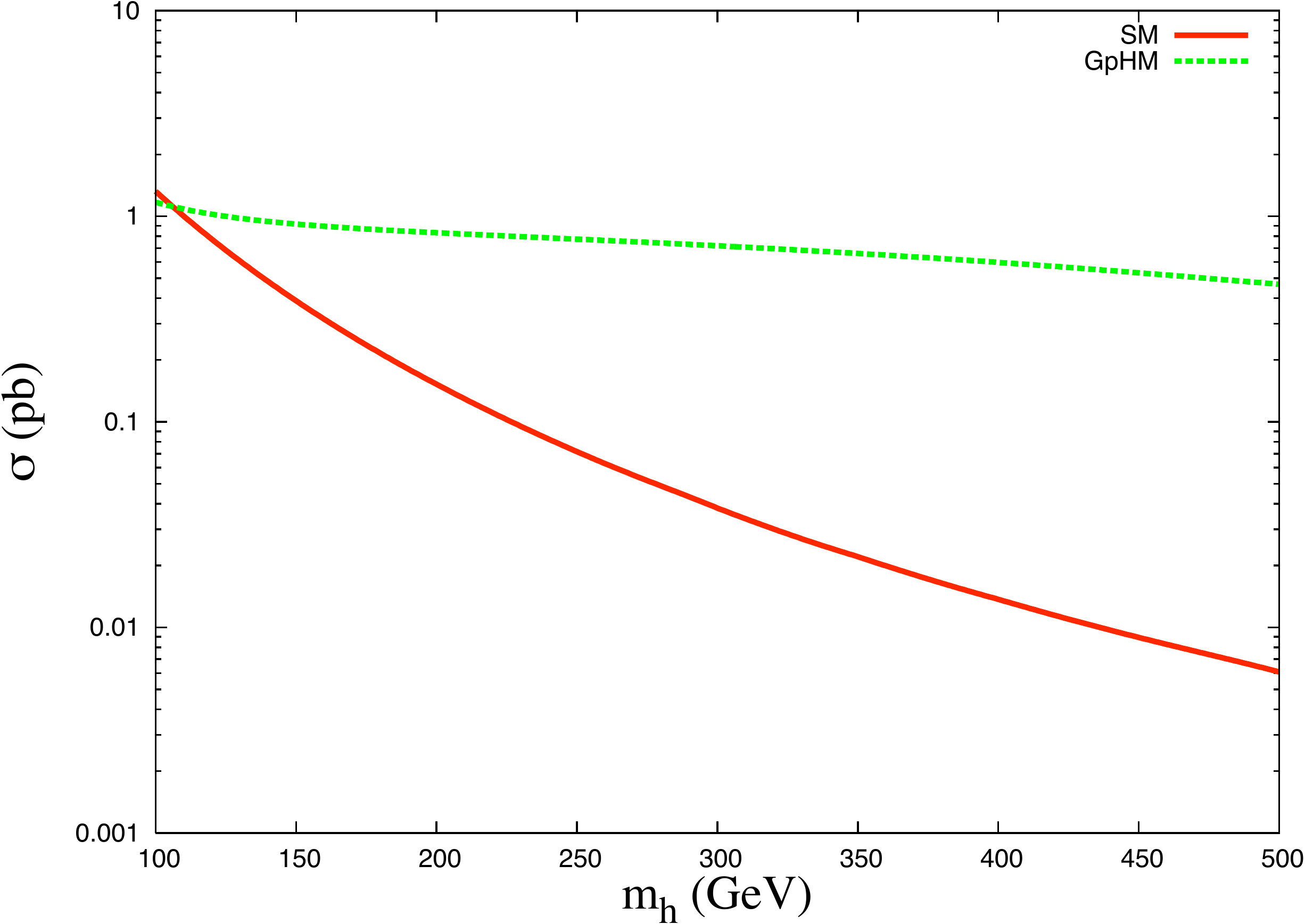}}
\caption{Cross-section in $pp \to WH$ through a $W'$ for a  range of Higgs masses.  }
\label{fig:xi_WHb}
\end{figure}

The production of a Higgs via radiation from a $W'$ is clearly important in the examination of this sort of model, where both KK states {\it and} a Higgs are present.  It is modes like this that can singlehandedly distinguish this model from extra-dimensional Higgsless models as well as from the SM.  This signal is ideal even for a light ($\sim 120$ GeV) Higgs, the benefit coming from the fact that we can now minimize background with a cut on the lepton coming from the decay of the $W$.  In Fig.~\ref{fig:xi_WHb} we see that the $W^{(2)}WH$ coupling  increases the cross-section $\sigma(pp \to WH)$ at all masses $m_H \gtrsim 110\ {\rm GeV}$.  It is interesting to note results from the Tevatron \cite{D0WH} limiting this cross-section at $\sqrt s = 1.96 \, {\rm TeV}$: for a $ 115 \, {\rm GeV}$ Higgs mass an upper bound $\sigma (pp\to WH) \lesssim 11.4 \times  \sigma_{\rm SM}(pp \to WH)$ is determined.  For $m_H \sim 150 \, {\rm GeV}$ the upper bound already exceeds $50 \times \sigma_{\rm SM}(pp \to WH)$ so that no constraints on the GpHM can be inferred.

\begin{figure}[htb]
\centerline{\includegraphics[width=11cm]{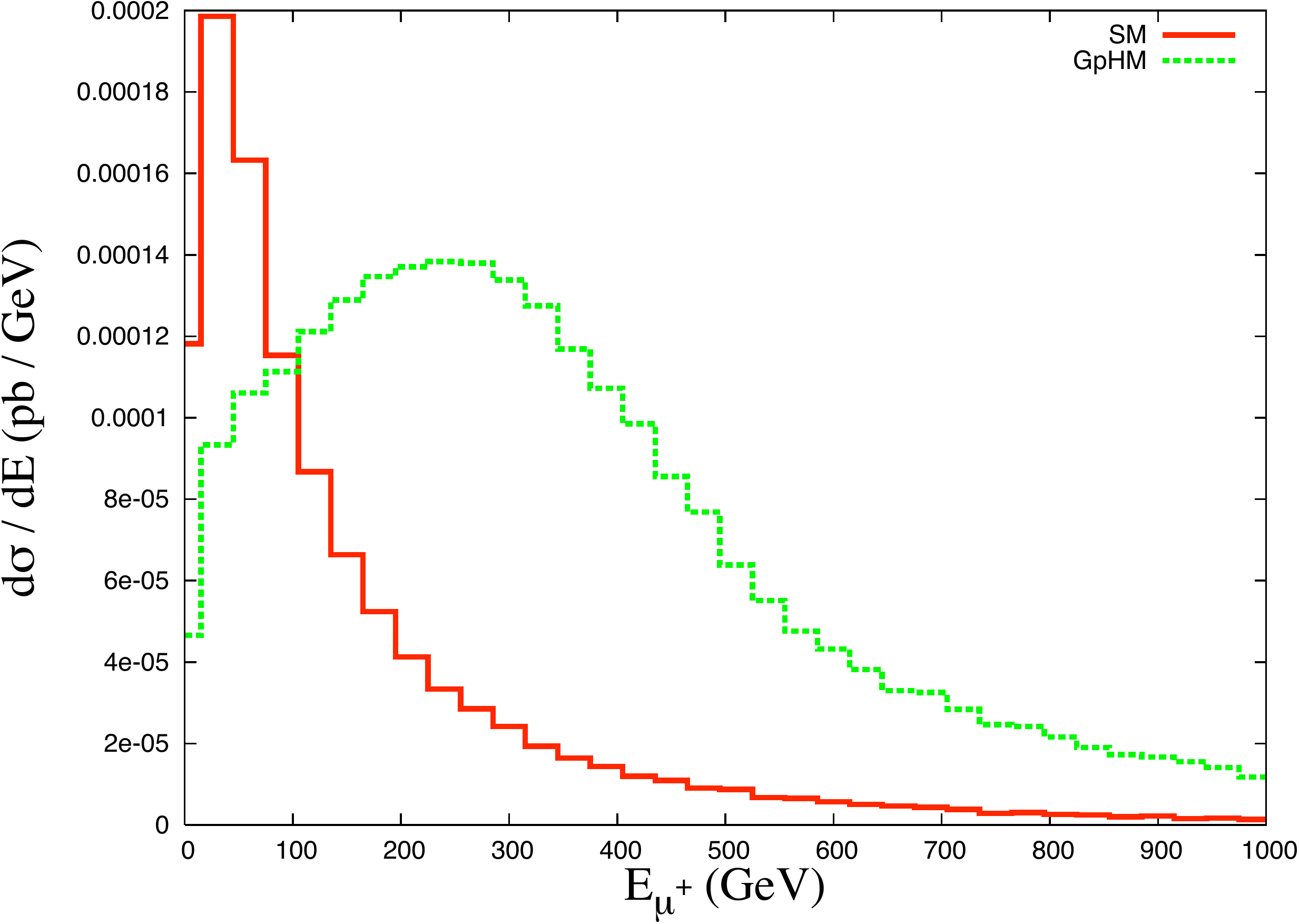}}
\caption{The gaugephobic Golden Channel $WH$; in this case $W^+$ decaying to $\mu^+ \nu$. The energy distribution of a final state isolated muon is plotted assuming  a Higgs with mass of $160$ GeV.  The plot shows the distribution coming from $pp \to W^+H \to \mu^+ \nu b\bar b$ without cuts, illustrating the reducibility of the SM background.}\label{fig:Emu}
\end{figure}
\begin{figure}[ht]
\centerline{\includegraphics[width=11cm]{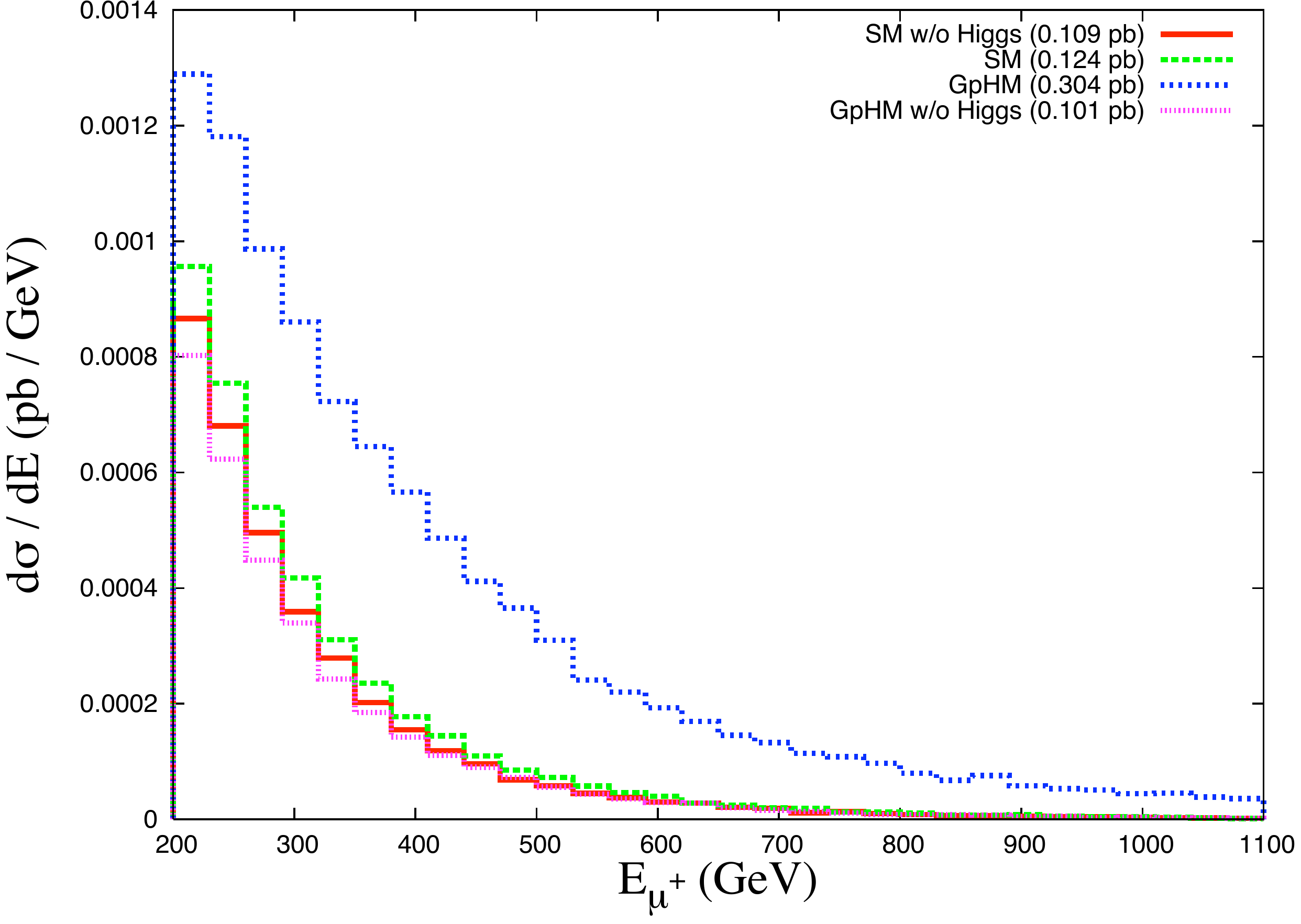}}
\caption{The distribution through all channels for $pp \to W^+ H \to \mu^+\nu b \bar b$ with a required minimum energy of 200 GeV taking $|\eta_\mu|,|\eta_b|<2.5$ and $m_{b\bar b}>130 \, {\rm GeV}$.}\label{fig:Emub}
\end{figure}

The dominant SM backgrounds to this channel come from $Wb\bar b$, $t\bar t$ and single top production 
\cite{D0WH,singletop}.
To illustrate the reducibility of this background we show the lepton (specifically muon) energy distribution in the gaugephobic scenario as well as  the SM in Fig.~\ref{fig:Emu}.  Supposing the Higgs decays to bottoms, we can minimize SM contributions where the number of events of a single  isolated hard lepton with a $b \bar b$ pair is small.  The energy distribution of a final state isolated muon after a minimum energy cut has been made is shown compared to background in Fig.~\ref{fig:Emub} as an illustration of this.  Thus with these cuts and 1 fb$^{-1}$ of data we would expect 304 signal events versus 124 SM background events. Note that with the Higgs artificially excluded, the GpHM predictions fall slightly below SM background, reflecting the fact that $g_{Wtb}/g_{Wtb{\rm (SM)}} \approx 0.96$ at this benchmark.

\begin{figure}[hc]
\centerline{\includegraphics[width=11cm]{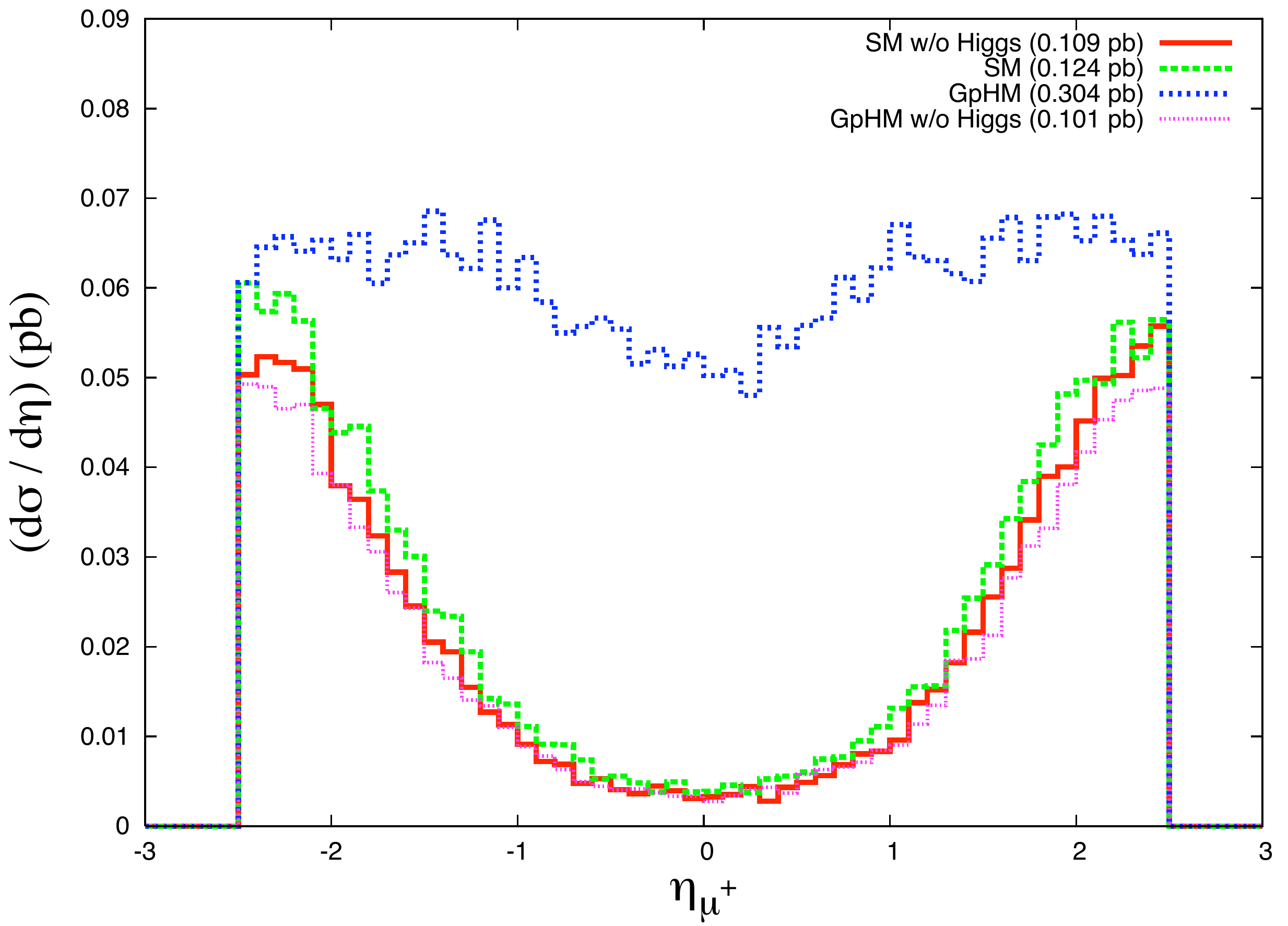}}
\caption{Distribution of $\mu$ pseudorapidities in $pp \to \mu^+ \nu b \bar b$.  The contribution of new heavy gauge bosons increases the central peak.}
\label{fig:Eta}
\end{figure}

\begin{figure}[htb]
\centerline{\includegraphics[width=11cm]{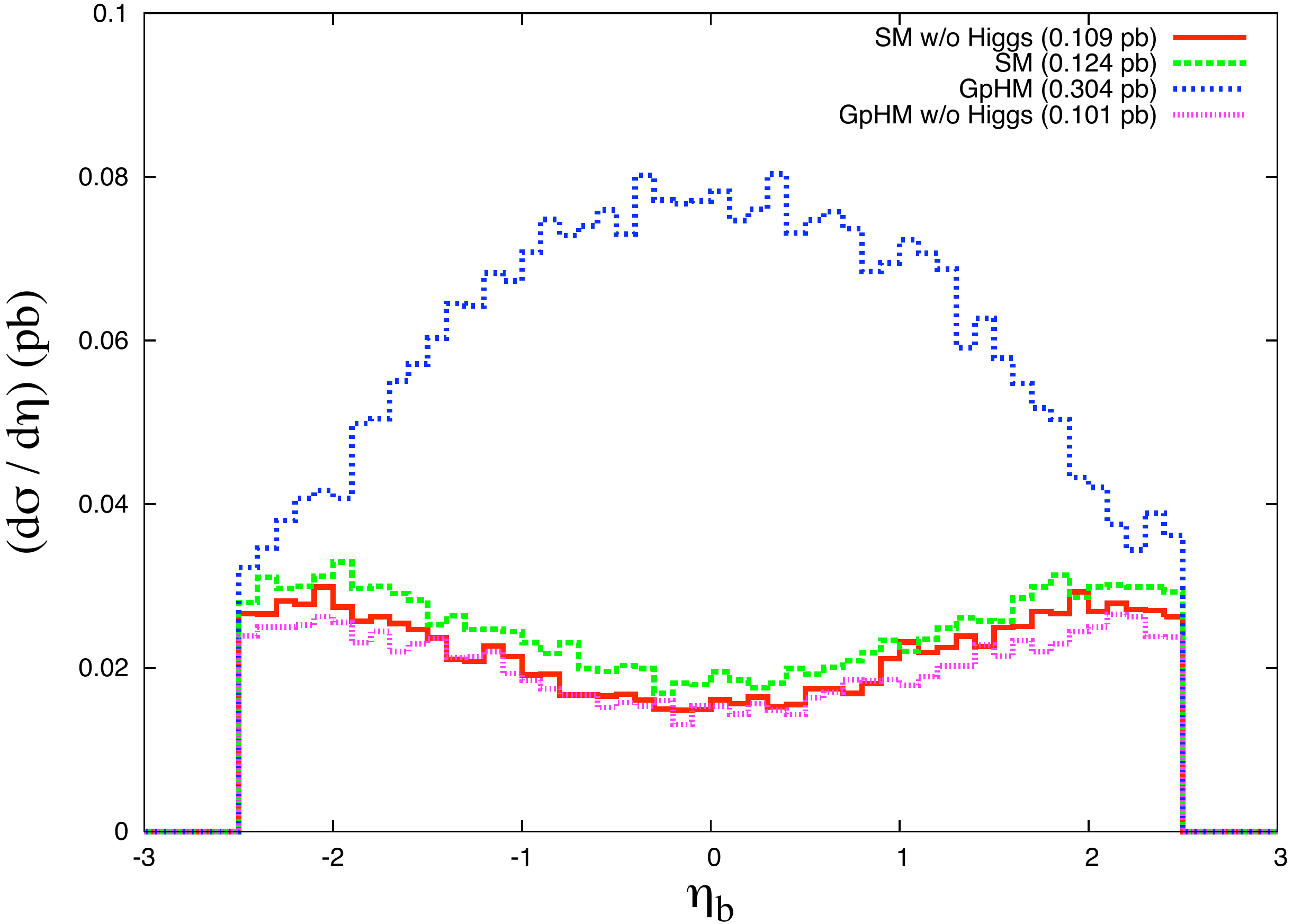}}
\caption{Distribution of $b$ quark  pseudorapidities in $pp \to \ell {\bar \nu} b \bar b$.  The contribution of new heavy gauge bosons increases the central peak.}
\label{fig:Etab}
\end{figure}
The pseudorapidity  of the decay products has a more central distribution than in the SM, and this provides a particularly clear signal of heavy vector bosons.  The $\eta$ distributions for the final state muons and bottom quarks are shown in Figs.~\ref{fig:Eta} and ~\ref{fig:Etab}.

\subsection{Additional Probes in GpHM}
In the previous section we exploited the strength of the $W^{(2)}WH$ coupling in Higgsstrahlung processes.  There are other instances of enhancement coming from processes involving also the participation of new colored fermion states; we mention the principal ones in Table~\ref{table:Processes}.  

\begin{table}[htb]
\begin{center}
\begin{tabular}{c|c|c}
Process & $\sigma_{\rm SM}$ (fb)  & $\sigma_{\rm GpHM}$ (fb)\\
\hline \hline
$pp \to 2H +2j$ & $\sim 1$ &27 \\
$pp \to t \bar t $ & $1139 \times 10^3$ & $1239 \times 10^3$ \\
$pp \to t \bar t + 2 W$  & 94.8 & $1.05 \times 10^3$ \\
$pp \to t \bar t +2Z$ & 4.6 & 326 \\
$pp \to t \bar t +2Z+H$ & 0.03 & .98  \\ 
$pp \to W + H$ & 188 & 403\\
\hline 
\end{tabular}\caption{$pp$ processes (after minimal cuts) with enhancement from exotic quark contributions in GpHM at the Mixed gaugephobic benchmark point with $V = 300$ GeV.  Here $M_H = 160$ GeV.}\label{table:Processes}
\end{center}
\end{table}

In our study, we've focused primarily on the contributions to TeV-scale physics coming from the interplay of a gaugephobic Higgs and KK weak gauge bosons.  However, we can clearly see in the cross-sections given in Table~\ref{table:Processes} that the new quarks can have dramatic impacts.  As described in Section~2, there is a necessary custodial protection of the coupling $g_{Zb \bar b}$ that suggests the introduction of the quarks $X$ and $T$.  In the processes involving $t\bar t$ pairs the effects of these new quarks are seen to be significant.  The rates in Table~\ref{table:Processes} are inclusive in that they sum over all intermediate KK resonances, $X$, and $T$ states which can lead to the indicated final states.

As we are interested in differentiating Gaugephobic from Higgsless and the Standard Model, we don't pursue the large exotic quark production further here.  A study involving generically similar heavy quark states can be found in \cite{Contino}.  It is important to note that the requirements imposed by constraints on $Z \to b \bar b$ can lead to additional distinct model signatures.   

What would be interesting to distinguish the Gaugephobic model from Higgsless and the SM would be to see Higgs bosons coupling to these extra quarks $X$ and $T$.  First it is clear that these signals are several orders of magnitude smaller than the dominant and clean $W^\pm H$ signal.  The interesting signals for differentiation involve the final states $t \bar t +2Z+H$ and $t \bar t+2W+H$.  These are extremely high multiplicity final states, and arise with similar cross sections to $WH$ because while the QCD production enhances the cross section, the phase space is reduced by the necessity to pair produce the $X$ or $T$.  Allowing two of the $W$'s to decay to leptons, the detector level objects are 4 $b$-jets, 4 light jets, 2 leptons plus missing energy.  Such events should be easy to trigger on because the total energy is very large, and there's a possibility for a hard isolated lepton as well.

Such an analysis is in principle possible but suffers from a substantial combinatorial background and overlap of jets.  Eight jets and two leptons will generally always have two jets or a jet and lepton close to each other in $R$.  After paying the price of branching fractions and efficiencies, there would only be a handful of events in each channel at the LHC with $300\ {\rm fb}^{-1}$.

We have not considered single production of $X$ and $T$ because this requires additional model input for the Yukawa couplings of these new states.  Furthermore, this Yukawa must be small, making pair production preferable.

\section{Conclusions}
\label{sec:conclusions} \setcounter{equation}{0}
\setcounter{footnote}{0}

The Gaugephobic Higgs model is an interesting model interpolating between RS1, the Higgsless model, and the Standard Model.  Differentiating it from these limiting cases will be an important task at the LHC, if new gauge bosons are found at the LHC, and/or Standard Model Higgs signals are suppressed or experimentally absent.

We have shown that due to the boundary conditions on the TeV brane and the resulting wave function profiles in the extra dimension, there is an enhanced coupling between the Higgs and even Kaluza-Klein states.  This happens simply because because the even modes can have an anti-node on the TeV brane.
This enhanced coupling  leads to  the clean signature in $W^\pm H$ which provides a powerful Higgs discovery mode at the LHC for any Higgs mass in this class of models.  We hope that experimentalists at ATLAS and CMS will be able to provide a more detailed analysis of this signal in the near future.

\section*{Acknowledgments}
We thank   Giacomo Cacciapaglia, Roberto Contino, Csaba Cs\'aki,  Cristoph Grojean, Guido Marandella, Tilman Plehn, and Andreas Weiler for useful conversations.  J.T. is
supported in part by the US Department of Energy under contract No.
DE- FG03-91ER40674 and by a US-Israeli BSF grant.

\end{document}